\documentclass[english,gc, manuscript]{copernicus}

\makeatletter

\providecommand{\tabularnewline}{\\}

\makeatother

\usepackage{babel}
\begin{document}
\title{Demonstrating change from a drop-in space soundscape exhibit by using
graffiti walls both before and after}
\author[1, 2]{Martin O. Archer}
\author[3]{Natt Day}
\author[3]{Sarah Barnes}
\affil[1]{Space and Atmospheric Physics, Department of Physics, Imperial College
London, London, UK}
\affil[2]{School of Physics and Astronomy, Queen Mary University of London,
London, UK}
\affil[3]{Centre for Public Engagement, Queen Mary University of London, London,
UK}
\correspondence{Martin O. Archer\\
(m.archer10@imperial.ac.uk)}
\runningtitle{Graffiti walls before and after a drop-in soundscape}
\runningauthor{Archer et al.}
\maketitle
\nolinenumbers
\begin{abstract}
Impact evaluation in public engagement necessarily requires measuring
change. However, this is extremely challenging for drop-in activities
due to their very nature. We present a novel method of impact evaluation
which integrates graffiti walls into the experience both before and
after the main drop-in activity. The activity in question was a soundscape
exhibit, where young families experienced the usually inaudible sounds
of near-Earth space in an immersive and accessible way. We apply two
analysis techniques to the captured before and after data --- quantitative
linguistics and thematic analysis. These analyses reveal significant
changes in participants' responses after the activity compared to
before, namely an increased diversity of language used to describe
space and altered conceptions of what space is like. The results demonstrate
that the soundscape was surprisingly effective at innately communicating
key aspects of the underlying science simply through the act of listening.
The impacts also highlight the power of sonification in stimulating
public engagement, which through reflection can lead to altered associations,
perceptions and understanding. Therefore, we show that this novel
approach to drop-in activity evaluation, using graffiti walls both
before and after the activity and applying rigorous analysis to this
data, has the power to capture change and thus short-term impact.
We suggest that commonly used evaluation tools suitable for drop-in
activities, such as graffiti walls, should be integrated both before
and after the main activity in general, rather than only using them
afterwards as is typically the case.
\end{abstract}

\introduction{}

Drop-in activities --- short, interactive, two-way engagements ---
tend to form a significant fraction of all non-school public engagement,
e.g. $31\pm3\%$ of all public activities across the UK's South East
Physics Network in 2017/2018 were less than 30~min in duration per
individual \citep{galliano18}. Such activities though are difficult
to effectively evaluate the impact of, since this necessitates a measure
of change on participants \citep{king15}. While surveys both before
and after may be one of the most robust methods of impact evaluation
in general \citep{jensen14}, these are neither appropriate for nor
commensurate with drop-in activities. This is because participants
are arriving all the time, the engagement duration is so short, and
surveys risk affecting participants' experience \citep{grand17}.
A number of evaluation tools more suitable for drop-in activities
have been reported including feedback cards, rating cards, snapshot
interviews, and graffiti walls \citep[e.g.][]{grand17,oxford19}.
Graffiti walls are large areas (often a wall, whiteboard, or large
piece of paper) where participants are free to write or draw responses
in reaction to the engagement activity or some prompt question, either
directly on the area itself or by sticking responses to it. All of
these evaluation methods for drop-ins are particularly useful in process
evaluation --- assessing the implementation of the activity. Under
typical usages (post-activity only) though, they are limited in their
ability to routinely demonstrate change from, and thus the impact
of, the engagement activity on participants in general.

This paper presents a novel implementation of graffiti walls for impact
evaluation, integrating them into both the start and end of a drop-in
activity. The activity was a soundscape experience surrounding current
space science research that used geostationary satellite data converted
into audible sound. We show that this evaluation method (through its
design, data collection, and analysis) can indeed capture immediate
impact --- changed language and conceptions of space in this case.
Appendices include details of statistical and qualitative coding techniques
employed throughout.

\section{Background\label{sec:Background}}

A common misconception is that space is a true vacuum completely devoid
of matter and thus there is no activity other than that of the celestial
bodies, e.g. planets or asteroids. However, the universe is permeated
by tenuous plasmas --- gases formed of electrically charged ions
and electrons that generate and interact with electromagnetic fields
\citep[e.g.][]{baumjohann}. One such example is the solar wind streaming
at several hundreds of kilometres a second from the Sun to the edge
of the heliosphere, something which only $58\pm2\%$ of the UK adult
population are aware of \citep{spaceweatherdialogue15}. Space plasmas
are also not just limited to our solar system, with other stars having
their own stellar winds \citep[e.g.][]{lamers99} and the interstellar
medium bridging the gap between these plasma bubbles in outer space
\citep{gurnett13}.

The presence of a medium in space allows for plasma wave analogues
to ordinary sound (pressure waves) that occur at ultra-low frequencies
--- fractions of milliHertz up to 1~Hz. They are routinely measured
by many space missions and can have perturbations that are significant
fractions of the background values. For a further discussion of the
equivalence of these plasma waves to sound see \citet{archer20entaudio}.
One way in which ultra-low frequency waves are generated is through
the highly dynamic solar wind buffetting against Earth's magnetic
field. This process plays a key role within space weather and thus
how phenomena from space can affect our everyday lives \citep[e.g.][]{Keiling2016}.
However, the belief by the public that space is completely empty in
turn leads many to incorrectly think that there is absolutely no sound
in space, reinforced by school science demonstrations such as the
bell-jar experiment (see \citealp{caleon13}, for a nuanced discussion
of this experiment and sound in near-vacuum conditions) or even popular
culture like in the marketing to the movie `Alien' (``in space no
one can hear you scream''). Public engagement with this research
area may help correct this fallacy.

Sonification --- the use of non-speech audio to convey information
or perceptualise data \citep{kramer94} --- can be used to convert
satellite measurements of these usually inaudible space sounds into
audible signals, simply by dramatically speeding up their playback
\citep{alexander11,alexander14}. This has already been leveraged
in public engagement projects for both scientific and artistic outputs
\citep{archer18,archer_ssfx}. Sonification in general has been applied
to various scientific datasets \citep{feder12}. \citet{supper14}
posits that through the public experiencing data in this way it can
grip their imagination and produce sublime experiences because of
sound's immersive and emotional nature. These arguments, however,
are mostly based on reflections from researchers and artists, rather
than through the evaluation of participants' own thoughts and feelings.
This paper evaluates the short-term impact on participants of experiencing
the sounds of space using graffiti walls both before and after a soundscape.

\section{Space Soundscape Exhibit\label{sec:Space-Soundscape-Exhibit}}

\begin{figure*}
\begin{centering}
\includegraphics[width=0.8\columnwidth]{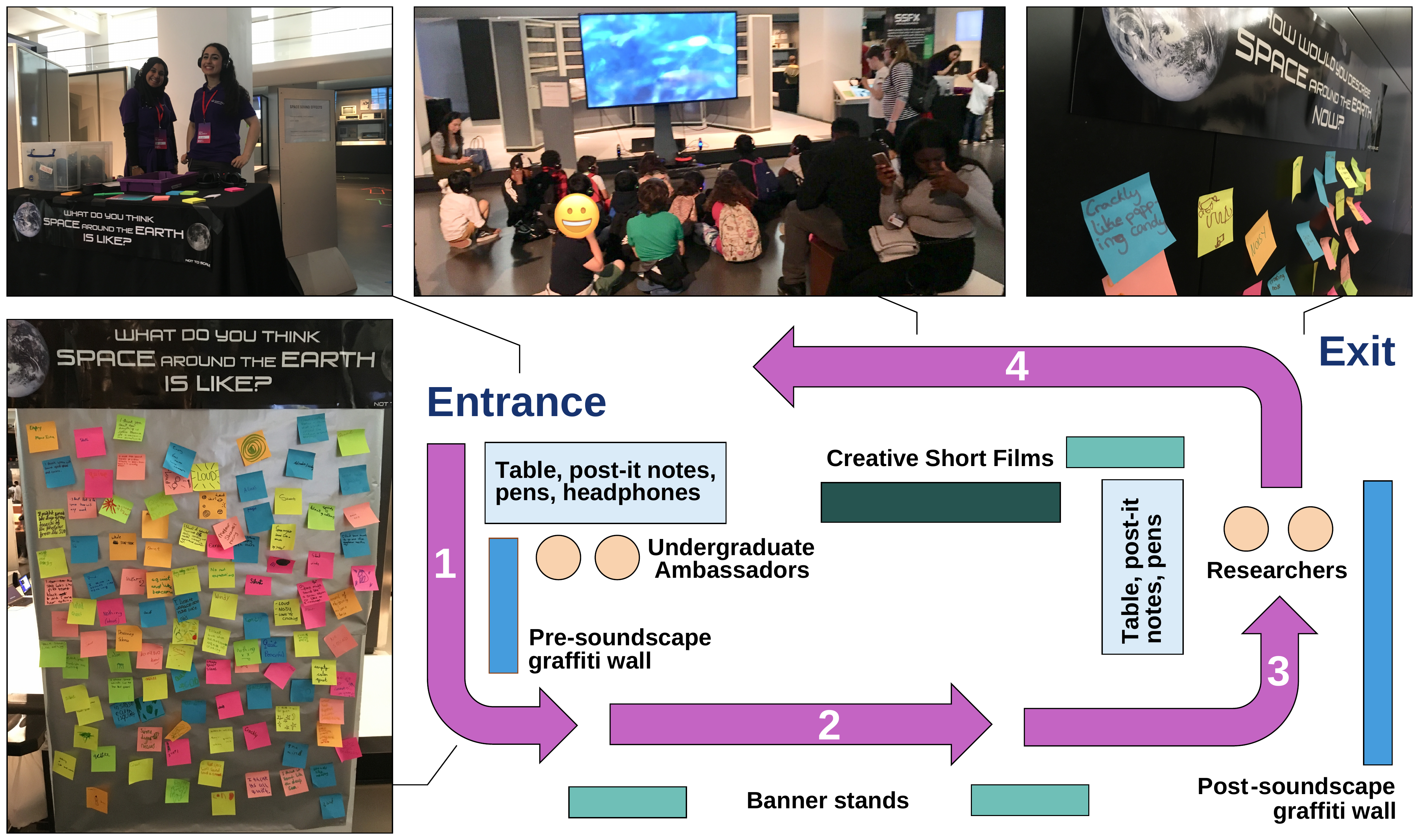}
\par\end{centering}
\caption{Layout and photos of the soundscape exhibit.\label{fig:layout}}
\end{figure*}

The space soundscape exhibit was held at the free Science Museum in
London (United Kingdom) whose informal learning adopts an inclusive,
accessible `science capital' approach that attracts a diverse range
of audiences \citep{smg17,smg20}. `Science capital' is defined as
the total science-related knowledge, attitudes, experiences, and resources
that a person has built up over their life \citep{archer17}. This
includes what science they know about, what they think and feel about
science, the people they know and their relation to science, and the
day-to-day engagement they have with science. The exhibit formed part
of the museum's `Summer of Space Season', held in celebration of the
50th anniversary of the Apollo moon landings, for which the museum
solicited drop-in space-themed activities aimed at young families.
It ran between the hours 12:00--16:00 during the May 2019 half-term
school holiday over the course of 4~days.

The purpose of the space soundscape was primarily to provide young
children and their parents/carers (as key influences upon them) an
accessible and immersive experience with space research that would
enable participation and spark discussion. Such experiences may, when
taken in conjunction with all the other formal and informal interactions
with science afforded to a young person, contribute towards developing
their science identity and hence build their `science capital'. Using
a generic learning outcomes framework \citep{hoopergreen07}, the
main intentions for the activity fell within the realms of `Enjoyment,
Inspiration, Creativity' and `Attitudes \& Values', with explicitly
enhancing `Knowledge \& Understanding' being only a secondary aim.
Figure~\ref{fig:layout} shows the layout of the exhibit, which was
integrated amongst the museum's usual collections, along with accompanying
photos. The activity worked as follows:
\begin{enumerate}
\item Museum attendees were invited to participate at the entrance by undergraduate
ambassadors. They were first asked to write or draw on a post-it note
what they think space around our planet is like. Some younger children
required further prompting beyond this broad question however, with
ambassadors often asking ``what do you think space sounds like?''
The participants placed their responses on the pre-soundscape graffiti
wall and were handed bluetooth wireless headphones playing the sounds
of space.
\item Participants went on a journey while listening to the sounds, following
a set of coloured arrows marked out on the floor. A number of banner
stands with further information about the sounds were placed along
this path, though it was observed that few people read these. This
may be either because participants preferred to listen to the sounds
or that it was not clear the stands were part of the experience given
the exhibit's location amongst other collections.
\item Near the end of the journey, researchers took participants' headphones
and asked them to reflect on what they think about space after having
listened to the sounds. Participants then recorded their thoughts
on post-it notes again and placed these on the post-soundscape graffiti
wall. The researchers would use what they had written or drawn to
prompt a short dialogue about aspects of the space environment around
Earth and space weather research. This method was informed by the
`science capital' research \citep{archer17}, which recommends scientists
use and value participants' own experiences within their engagement
practice to help enable lower `science capital' audiences to feel
included in science and that science is for \textquotedblleft people
like me\textquotedblright . These discussions provided an opportunity
to solidify, or in some cases clarify, the associations that participants
made from the soundscape experience in a tailored and audience-focused
way (e.g. only going into an appropriate level of detail depending
on the individual or group).
\item Finally, researchers would change the channel on the headphones so
that participants could watch on a large TV screen a series of creative
short films inspired by and incorporating the sounds \citep{archer_ssfx}.
The films also featured epilogue text reinforcing the importance and
relevance of space weather research. Surprisingly, these artistic
films proved much more popular than anticipated.
\end{enumerate}
The graffiti walls were used as an open opportunity for participants
to reflect upon their perceptions and associations with space both
before and after the soundscape, with this being intentionally left
broad to elicit a wide range of possible responses and thus potential
impacts. This method was chosen specifically due to its suitability
for evaluating drop-in activities, ability to be integrated within
the activity itself, and alignment with our intended overall experience
for participants. While graffiti walls are a common evaluation tool,
we are unaware of any published public engagement activity that has
captured and analysed data both before and after a drop-in activity
using them. This makes our evaluation approach for the exhibit novel.

Ethical considerations in the design of the exhibit and its evaluation
followed the British Educational Research Association \citep{bera18}
guidelines and were discussed with institutional funders and the Science
Museum before the activity occurred. All respondents consented to
providing graffiti wall responses as these were not mandatory for
participation in the soundscape exhibit. Children only participated
in any of the activities when accompanied by their appropriate adult.
All data collected was anonymous and no characteristics about participants
were solicited. Overall it was deemed (due to the nature of the exhibit,
its design, and the types of responses being collected) there was
very little risk of harm arising from participation.

The space soundscape was experienced by 1,003 people, recorded using
a tally counter. The majority were in family groups (approximately
three-quarters were children based on observations) with some independent
adults also. It was observed that in families typically only the children
contributed to the graffiti walls (with no substantive difference
in respondents before and after) and in many cases accompanying adults
did not take headphones when offered, perceiving the activity as just
for their children. There were 535 and 446 responses (predominantly
textual) on the pre- and post-soundscape graffiti walls respectively,
rates of $53\pm2\%$ and $44\pm2\%$. This is some 3--10 times greater
than reported for typical graffiti walls \citep{oxford19} likely
due to their integration into the overall activity here.

\section{Results and Analysis}

\begin{figure*}
\centering{}\includegraphics[width=0.66\columnwidth]{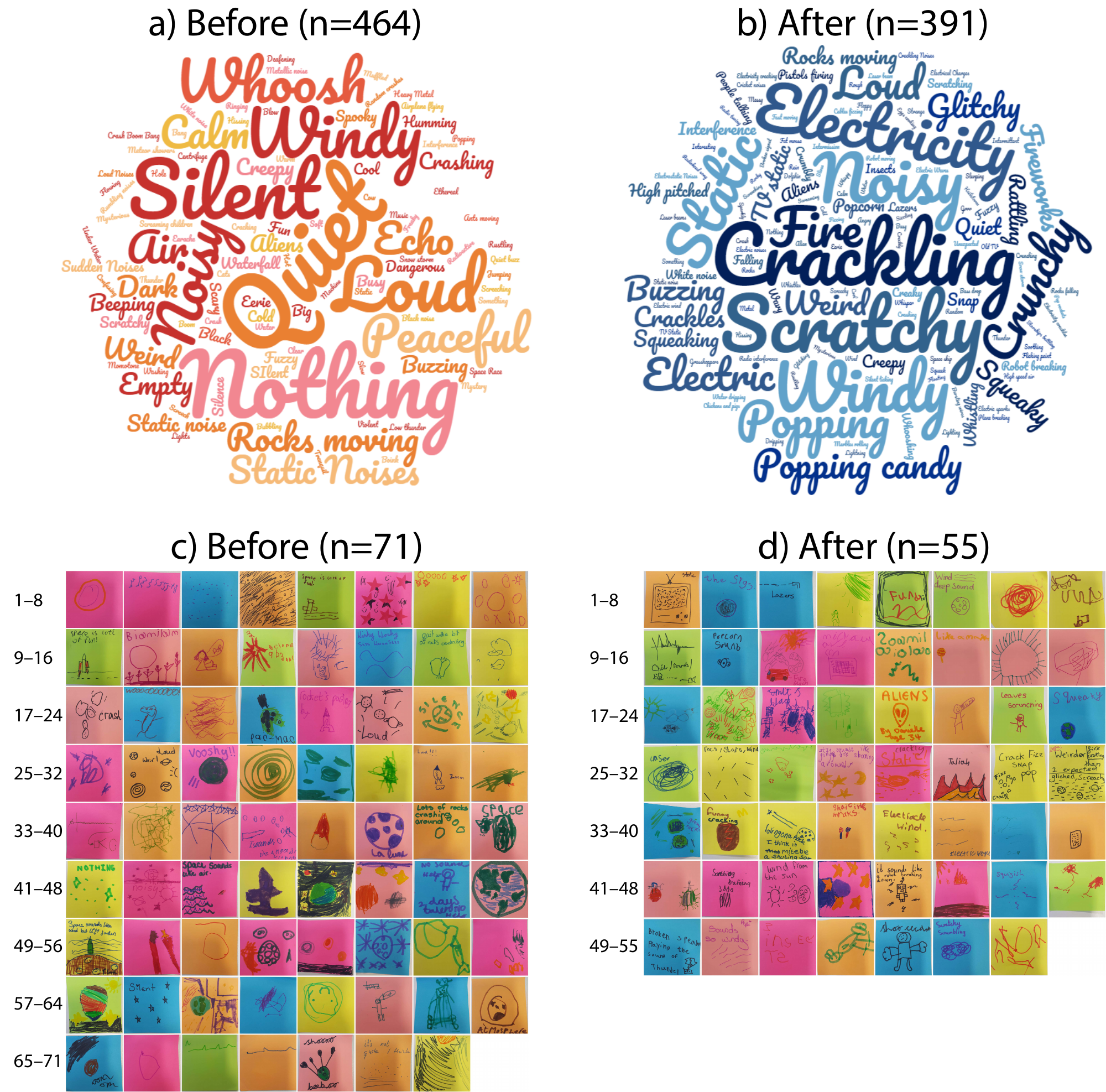}\caption{Wordclouds (a,b) and drawn images (c,d) from both before (a,c) and
after (b,d) experiencing the soundscape.\label{fig:before-after}}
\end{figure*}

The data captured on the pre- and post-soundscape graffiti walls are
displayed in Figure~\ref{fig:before-after}. However, simply presenting
these is insufficient to robustly demonstrate any potential changes
and thus impacts. Instead, analysis is required and two approaches
are taken here, namely quantitative linguistics and thematic analysis.

\subsection{Quantitative linguistics}

Quantitative linguistics investigates language using statistical methods
and has uncovered several linguistic laws that mathematically formulate
empirical properties of languages. One of these is Zipf's law ---
the frequency of words are approximately inversely proportional to
their rank (where the more often a word is used the higher its rank,
i.e. closer to 1) \citep{zipf35,zipf49}. An alternative way this
law is stated is that the statistical distribution of word ranks follows
a power law with an exponent that is typically quoted as $-1$ . Zipf's
law holds well for almost all languages as well as many other human-created
systems \citep{piantadosi14}. The Zipf exponent, however, can vary
and is a measure of the diversity of words. \citet{baixeries13} showed
that children's Zipf exponents become less-negative / shallower with
age, demonstrating increasing variety of language and thus linguistic
complexity as they develop. However, we are not aware of Zipf's law
being exploited in public engagement evaluation before.

\begin{figure}
\begin{centering}
\includegraphics{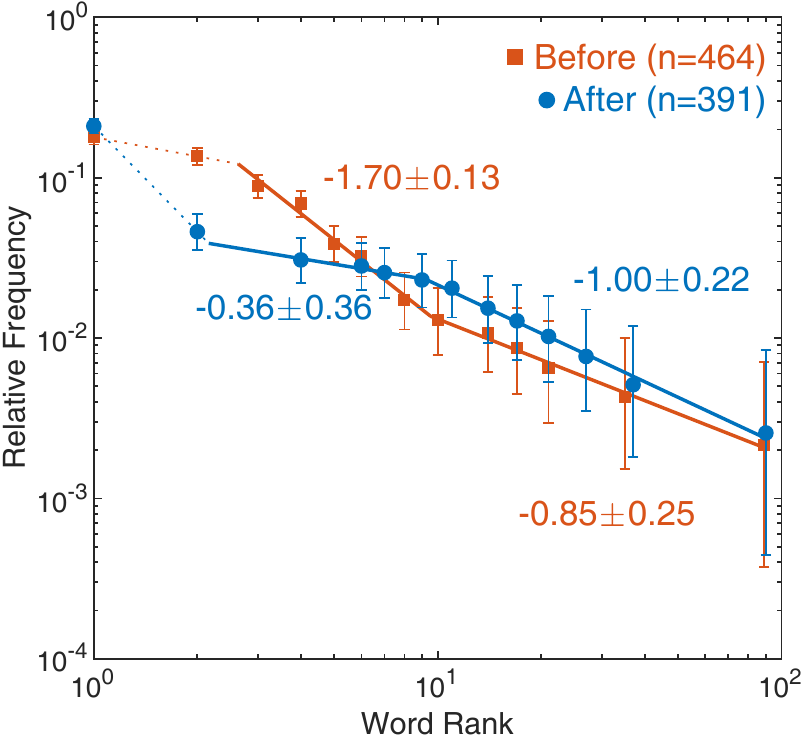}
\par\end{centering}
\caption{Log-log rank-frequency plot of words before (orange) and after (blue)
the soundscape. Power law exponents from a piecewise linear regression
are indicated. Uncertainties refer to standard errors.\label{fig:zipf}}
\end{figure}

Figure~\ref{fig:zipf} shows rank-frequency plots of the textual
responses to the soundscape before and after the experience. This
particular analysis thus omits any purely pictorial responses. Ties
in ranks have been accounted for using standard competition ranking
(also known as ``1224'' ranking, where a gap is left following the
tie). It is clear from these plots that the distributions follow broken
power laws (apart from the top word which is of similar frequency
before and after). Break points and exponents have been ascertained
by a piecewise regression (see Appendix~\ref{sec:stats}). Interestingly,
the breaks in the two datasets occur at similar ranks namely $\sim$2--3
and $\sim$9--10. We are not concerned with the specific values of
the Zipf exponents, which could depend on the demographics of participants,
but simply whether they changed from before to after and in what sense.
The exponents in the higher rank segments show clear differences ---
the after dataset exhibits a much shallower exponent. The lowest ranked
segments are, in contrast, consistent with one another. The top 10
ranks constitute $62\pm2\%$ of words before and $45\pm3\%$ after,
making the two entire distributions significantly different ($p=8\times10^{-11}$
in a two-sample Kolmogorov-Smirnov test, see Appendix~\ref{sec:stats}).
The overall effect is an increased diversity of words resulting following
the soundscape. We interpret this positive impact as signifying the
participants engaged with and reflected on the stimulating experience
afterwards, rather than continuing to draw from common associations
concerning space which they likely did beforehand. We have therefore
demonstrated language change in participants resulting from a public
engagement activity through the novel usage of Zipf's law applied
to graffiti wall responses.

\subsection{Thematic analysis}

Thematic analysis \citep{Braun2006} was used to analyse the meaning
behind both textual and drawn responses. This finds patterns, known
as qualitative codes, in the data which are then grouped into broader
related themes. Instead of using pre-determined codes, the analysis
drew on grounded theory \citep{Robson2011,Silverman2010}, allowing
the themes to emerge from the data itself as outlined in Appendix~\ref{sec:codes}.
This more exploratory and data-driven approach enables unexpected
outcomes and impacts (be they positive or negative) to come to light,
rather than analysing the qualitative data only through a particular
lens based on specific intended outcomes. The main themes and underlying
(typically antithetical) codes determined by the first author are
given in Table~\ref{tab:themes}.

\begin{table}
\begin{centering}
\begin{tabular}{ccl}
\textbf{Themes} & \textbf{Codes} & \textbf{Description}\tabularnewline
\hline 
\multirow{2}{*}{Sound} & Quiet & Space is ``silent'' or ``quiet''\tabularnewline
 & Loud & Space is ``loud'' or ``noisy''\tabularnewline
\hline 
\multirow{2}{*}{Emptiness} & Empty & Space is an ``empty'' vacuum with ``nothing'' in it\tabularnewline
 & Full & Space is filled with material or activity such as ``wind''\tabularnewline
\hline 
\multirow{2}{*}{Dynamism} & Slow & Space is slow (e.g. ``calm'' or ``peaceful'')\tabularnewline
 & Busy & Space is highly dynamic exhibiting busy movement\tabularnewline
\hline 
Electricity & Electrical & Expressions of electrical phenomena\tabularnewline
\hline 
Space Objects & Space Objects & Commonly known celestial bodies (planets, stars, meteors etc.) or
artificial spacecraft\tabularnewline
\hline 
\end{tabular}
\par\end{centering}
\caption{Themes and underlying qualitative codes in the thematic analysis.\label{tab:themes}}
\end{table}

\begin{figure}
\begin{centering}
\includegraphics{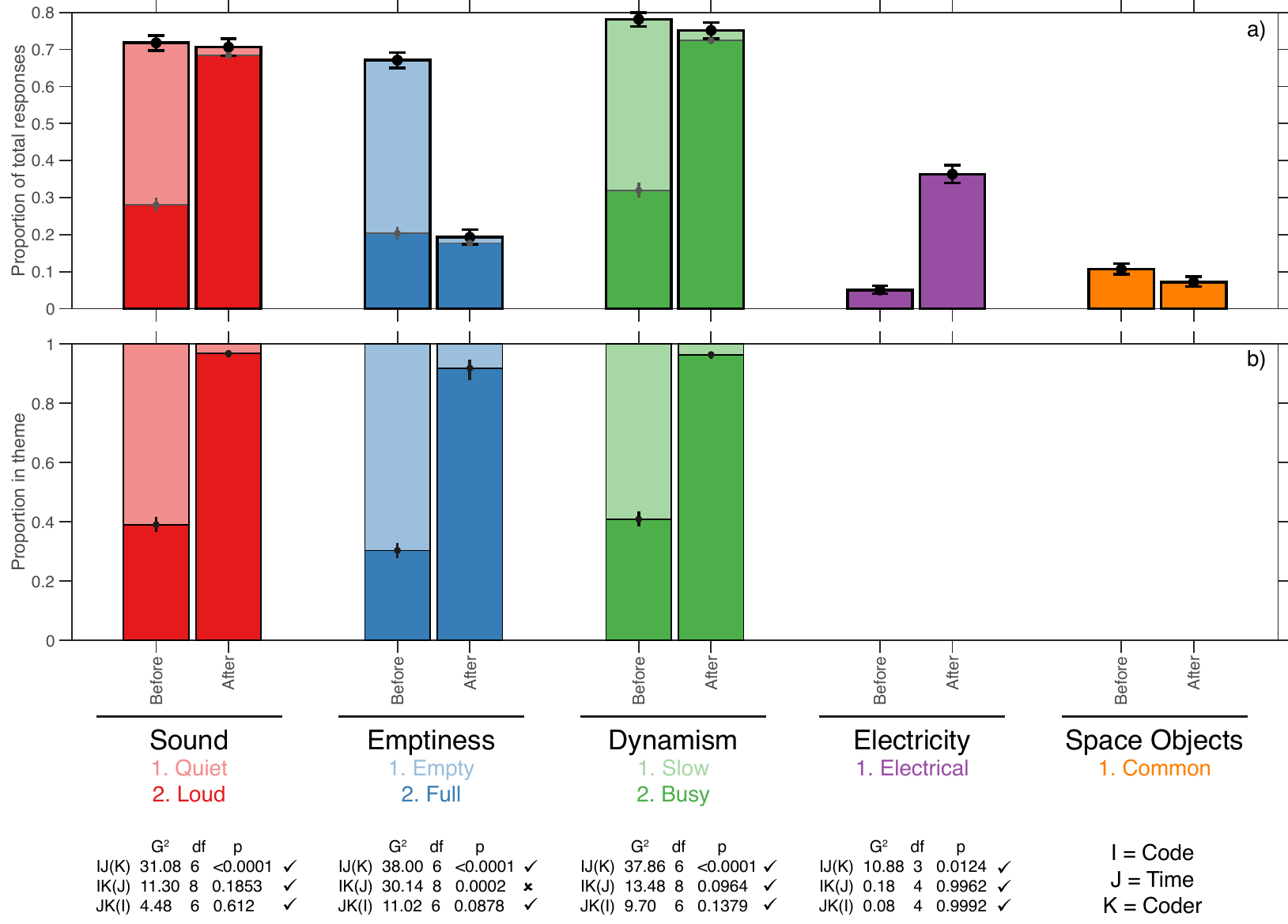}
\par\end{centering}
\caption{Comparison of qualitative themes and codes before ($n=535$) and after
($n=446$) the soundscape experience normalised by total responses
(a) and totals within each theme (b). Error bars depict the standard
error in proportions. Log-linear analysis statistics of the agreement
between coders are also shown for each theme.\label{fig:themes}}
\end{figure}

We quantify the number of responses in each theme and qualitative
code \citep[cf.][]{sandelowski01,sandelowski09,maxwell10} to investigate
any changes from before to after the soundscape experience. These
are shown in Figure~\ref{fig:themes} relative to the total responses
(panel~a) and within each theme (panel~b).

The theme of sound is highly relevant to the activity and was commonly
expressed both before and after. Responses beforehand mostly considered
space to be quiet/silent ($61\pm3\%$ within the theme). However,
a non-negligible fraction thought it to be loud, which may be due
to participants second-guessing the question because of the nature
of the activity and/or the phrasing by undergraduate ambassadors.
Nonetheless, the overwhelming majority ($97\pm1\%$ within the theme)
after the experience expressed space to be a noisy environment ---
a considerable change to beforehand. The perceived loudness of sound,
both in terms of human hearing and measurement, necessitates logarithmic
scales \citep{robinson56}. Such scales, like the decibel, therefore
require some reference base-level. For sound this is typically set
at the threshold pressure for human hearing of $20\,\mathrm{\mu Pa}$
\citep{roeser}. One must remember though that pressure fluctuations
depend on the background pressure level also ($100,000\,\mathrm{Pa}$
at sea level). Therefore, while the absolute amplitude of variations
in space are clearly small, relative to the background they are large
(as was noted in section~\ref{sec:Background}) and thus one can
consider space to be ``noisy'' in this sense. Another equally valid
perspective is that the process of sonification has revealed the presence
of sound that would otherwise not be audible and thus participants
have discovered, thanks to the exhibit, that space is ``noisier''
than they had previously imagined.

We note that the theme of dynamism exhibits quantitatively similar
results to that of sound --- a clear majority ($59\pm3\%$ within
the theme) thought space to be slow beforehand, whereas the vast majority
($96\pm1\%$) consider it highly dynamic afterwards. The dynamism
of Earth's magnetosphere is relative to the natural timescales of
the system. The typical periods of oscillations are of the order of
several to tens of minutes, and the properties of the waves (and even
their drivers) can significantly change within just a few wave periods
\citep[e.g.][]{Keiling2016}. This is unlike most sounds we are used
to on Earth, which often remain coherent for many hundreds or even
thousands of oscillations. Therefore, just like with sound, space
around our planet can be considered dynamic both relative to the properties
of the environment and relative to participants' prior expectations.

The theme of emptiness (including both of its underlying codes) was
quite common in responses beforehand, however it was expressed much
less often following the soundscape. The prevailing opinion before
was that space is empty and this dramatically reduced following the
soundscape, both relative to the total responses (from $47\pm2\%$
to $2\pm1\%$ ) and within the theme (from $70\pm3\%$ to $8\pm4\%$).
In contrast, the expression of space being full was communicated a
similar number of times both before and after. Therefore, participants
that had previously thought space was empty typically went on to write
words that fell within a different theme, rather than a response signifying
space as being filled with material. Since space is not absolutely
devoid of material, being permeated by tenuous plasmas, the exhibit
has successfully challenged this common misconception.

There was a clear increase in the proportion of responses relating
to electricity following the event, from $5\pm1\%$ to $36\pm2\%$.
Electricity is of fundamental importance to the plasma state, and
thus the increased realisation of this by participants is a welcome
change resulting from the exhibit.

Common space objects such as planets, stars, or satellites (typically
expressed through drawings) may appear at first glance of Figure~\ref{fig:before-after}
to be more frequent before the soundscape than after. As a fraction
of the total number of responses though, this difference is small
and not strictly statistically significant ($p=0.057$).

We checked the reliability of all these trends resulting from the
qualitative coding by applying log-linear analysis to a subset of
the data additionally coded by the co-authors (see appendices for
details). Using the notation that $I$ denotes the qualitative codes,
$J$ the time (i.e. before or after), and $K$ the different coders,
for the results to be consistent one would expect that the $IJ(K)$
test be statistically significant, constituting the reported trends
in codes with time, but the $IK(J)$ and $JK(I)$ interactions should
not be, indicating independence from individual coders. These statistics
are displayed in Figure~\ref{fig:themes} for each theme (apart from
space objects which was less common) indicating the expected behaviour
apart from in the case of emptiness. This theme showed some inconsistency
between coders for the ``full'' code, whereas when only ``empty''
was considered coders were in agreement ($G^{2}=32.2,3.42,2.06$ respectively).
Therefore, the main results of the paper are robust and hence we have
demonstrated a change in conceptions of space, well-aligned with the
underpinning research, that resulted from this drop-in engagement
activity.

\conclusions{}

A challenge within public engagement is evaluating the impact of drop-in
activities since this necessitates a measure of change using evaluative
tools that are appropriate to and commensurate with the engagement
\citep{jensen14,king15,grand17}. We have presented a novel implementation
and analysis stemming from a common evaluation tool, graffiti walls
\citep[e.g.][]{oxford19}. These were integrated both before and after
a soundscape exhibit on space science research using sonified satellite
data. The pre- and post-soundscape graffiti walls provided data on
participants' conceptions of space and, through their integration
into the activity itself, had much higher response rates than is typical.
The captured data was analysed in two different ways.

We investigated the statistical properties of the words expressed
by using Zipf's law from quantitative linguistics. This states that
the frequency of words in languages typically follow power laws whose
exponents give a measure of the diversity of words, where shallower
exponents indicate greater variety. The distributions from the graffiti
walls showed that the exponent for the top $\sim$10 words (constituting
$62\pm2\%$ of the responses before and $45\pm3\%$ after) became
significantly shallower from before to after, whereas the exponents
were consistent for the remaining words. This demonstrates an overall
increased linguistic complexity concerning participants' thoughts
about space following the activity. This positive result aligns with
the exhibit's aims in the realm of `Enjoyment, Inspiration, Creativity'
\citep[cf.][]{hoopergreen07}, since being exposed to the sounds of
space led to stimulation, reflection, and ultimately a more diverse
and creative set of words about space than had been expressed beforehand.
We are unaware of Zipf's law being used in impact evaluation for public
engagement before.

We also investigated themes present in the responses, which again
yielded significant and robust positive changes from before to after.
Beforehand participants typically expressed common misconceptions
of space being completely empty, silent, and with little activity.
However, after experiencing the space sounds they felt space was a
noisy and dynamic environment with electrical phenomena present. It
is astounding that simply by listening to the sounds these simple
aspects of the underlying space plasma physics were successfully and
innately communicated to participants before they even spoke to the
researchers. This therefore demonstrates the power of sonification
for audiences. While this had been argued by \citet{supper14} based
on reflections from researchers and artists, here we have shown it
from evaluating participants' experiences directly. Therefore, we
have shown postive effects in the realms of `Knowledge \& Understanding'
and `Attitudes \& Values' \citep[cf.][]{hoopergreen07} resulting
from the soundscape. The measured changes in associations, conceptions,
and perceptions will have been further reinforced by researchers drawing
from participants' own reflections in their subsequent dialogues \citep[cf.][]{archer17}.

Overall, integrating existing evaluation tools suitable for drop-in
engagement activities, such as graffiti walls, both before and after
a drop-in activity can enable practitioners to demonstrate changes
resulting from the engagement and therefore its short-term impact.
However, typically such tools are only used following activities,
which limits the ability to demonstrate some measure of change and
thus impact. We suggest that our approach, both in terms of data capture
and analysis, should be adopted more regularly, not just for soundscape
exhibits, but for a range of different drop-in activities in general.

\appendix

\section{Statistical techniques\label{sec:stats}}

Statistical uncertainties in proportions are estimated using the \citet{clopper34}
conservative method based on the binomial distribution, where standard
(68\%) errors are shown throughout.

A piecewise linear regression in log-log space was used to minimise
the sum of squared error between the data and a model made up of a
specified number of line segments whose break points could be varied
iteratively. This was performed for an increasing number of segments,
each time calculating the degrees-of-freedom-adjusted $R^{2}$ which
accounts for the number of explanatory variables added to the model:
\begin{equation}
\overline{R}^{2}=1-\left(1-R^{2}\right)\frac{n-1}{n-m-1}
\end{equation}
where $R^{2}$ is the usual coefficient of determination, $n$ is
the number samples, and $m=2s-1$ is the total number of explanatory
variables in the piecewise linear model with $s$ segments. The final
model was selected as the first peak in $\overline{R}^{2}$ with $s$.
Any segments with only two datapoints are later ignored. The statistical
significance of the slopes was determined by ANCOVA with a multiple
comparison procedure \citep{hochberg}. The standard errors in the
slopes quoted are derived from a propagation of uncertainty in the
proportions within the linear regression.

A two-sample Kolmogorov-Smironov test is used to non-parametrically
test the equality of two probability distributions. It quantifies
the distance between two one-dimensional empirical (cumulative) distribution
functions $F_{1,n}\left(x\right)$ and $F_{2,m}\left(x\right)$ as
\begin{equation}
D_{n,m}=\sup_{x}\left|F_{1,n}\left(x\right)-F_{2,m}\left(x\right)\right|
\end{equation}
 where $\sup$ is the supremum function \citep{massey51}. The critical
value of this statistic is given by $\sqrt{-\frac{1}{2}\ln\left(\alpha/2\right)\left(m+n\right)/mn}$
for desired significance $\alpha$.

Finally, log-linear analysis is employed to check the consistency
of the changes in coding with time across the different coders. This
extension of the $\chi^{2}$ test of independence to higher dimensions
uses a similarly distributed statistic, the deviance, given by
\begin{equation}
G^{2}=2\sum O_{ijk}\ln\frac{O_{ijk}}{E_{ijk}}
\end{equation}
for observed $O_{ijk}$ and expected $E_{ijk}$ frequencies \citep{agresti07}.
Here we assess conditionally independent models denoted $IJ(K)$,
which tests the two-way $IJ$ interaction with the effects of the
$IK$ and $JK$ interactions removed. Computationally this calculates
$G^{2}$ for each level of $K$ summing the results, with $G^{2}$
having $(n_{I}-1)(n_{J}-1)n_{K}$ degrees of freedom.

\section{Qualitative coding\label{sec:codes}}

The qualitative coding process of thematic analysis drawn from grounded
theory involved the following steps:
\begin{enumerate}
\item Familiarisation: Responses (Figure~\ref{fig:before-after}) are studied
and initial thoughts noted.
\item Induction: Initial codes are generated based on review of the data.
\item Thematic Review: Codes are grouped together into themes and applied
to the full data set.
\item Reliability: Codes are applied to a subset of data by second coders
to check reliability of results.
\item Finalisation: Theoretical interpretation and narrative are formulated
from final coding.
\end{enumerate}
Table~\ref{tab:counts} shows the number of responses (both unique
and total) across words and pictures in each theme and its underlying
codes both before and after the soundscape experience. To ensure the
reliability of the main qualitative coding of the entire dataset,
second coders applied the thematic analysis to a subset of the data.
This subset constituted the top 16 words before (58\% of total responses)
and 15 words after (49\%), with the slightly different number of words
used in the two datasets being due to ties in the ranking of words
making it impossible to have exactly the same number in both. Table~\ref{tab:coders}
shows the totals of how these unique words were grouped across all
three coders. These results are used in the log-linear analysis to
test reliability, which we note does not require equally sized datasets.
The codes' association to the raw data can be found in the supplementary
material, both for the main and second coders.

\begin{table}
\begin{centering}
\begin{tabular}{llcccc}
\multirow{3}{*}{\textbf{Theme}} & \multirow{3}{*}{\textbf{Codes}} & \multicolumn{2}{c}{\textbf{Before}} & \multicolumn{2}{c}{\textbf{After}}\tabularnewline
 &  & Unique & Total & Unique & Total\tabularnewline
 &  & ($n=202$) & ($n=535$) & ($n=190$) & ($n=446$)\tabularnewline
\hline 
\multirow{3}{*}{Sound} & 1. Quiet & 22 & 234 & 7 & 10\tabularnewline
 & 2. Loud & 59 & 150 & 91 & 305\tabularnewline
 & Total & 81 & 384 & 98 & 315\tabularnewline
\hline 
\multirow{3}{*}{Emptiness} & 1. Empty & 36 & 250 & 4 & 7\tabularnewline
 & 2. Full & 31 & 109 & 37 & 79\tabularnewline
 & Total & 67 & 359 & 41 & 86\tabularnewline
\hline 
\multirow{3}{*}{Dynamism} & 1. Slow & 33 & 247 & 8 & 12\tabularnewline
 & 2. Busy & 74 & 171 & 111 & 323\tabularnewline
 & Total & 107 & 418 & 119 & 335\tabularnewline
\hline 
Electricity & 1. Electrical & 13 & 27 & 36 & 162\tabularnewline
\hline 
Space Objects & 1. Common & 51 & 57 & 154 & 32\tabularnewline
\hline 
\end{tabular}
\par\end{centering}
\caption{Number of responses (both unique and total) in each theme before and
after the soundscape.\label{tab:counts}}
\end{table}

\begin{table}
\begin{centering}
\begin{tabular}{ll|cc|cc|cc|}
 &  & \multicolumn{2}{c|}{Coder~1} & \multicolumn{2}{c|}{Coder~2} & \multicolumn{2}{c|}{Coder~3}\tabularnewline
\cline{3-8} \cline{4-8} \cline{5-8} \cline{6-8} \cline{7-8} \cline{8-8} 
 &  & Before & After & Before & After & Before & After\tabularnewline
\hline 
\multirow{3}{*}{Sound} & 1 Quiet & 8 & 0 & 5 & 0 & 8 & 1\tabularnewline
 & 2 Loud & 6 & 11 & 5 & 11 & 4 & 5\tabularnewline
 & None & 2 & 4 & 6 & 4 & 4 & 9\tabularnewline
\hline 
\multirow{3}{*}{Emptiness} & 1 Empty & 8 & 0 & 6 & 0 & 9 & 1\tabularnewline
 & 2 Full & 5 & 3 & 0 & 1 & 6 & 7\tabularnewline
 & None & 3 & 12 & 10 & 14 & 1 & 7\tabularnewline
\hline 
\multirow{3}{*}{Dynamism} & 1 Slow & 7 & 0 & 5 & 0 & 6 & 0\tabularnewline
 & 2 Busy & 7 & 11 & 4 & 12 & 10 & 11\tabularnewline
 & None & 2 & 4 & 7 & 3 & 0 & 4\tabularnewline
\hline 
\multirow{2}{*}{Electricity} & 1 Electrical & 2 & 6 & 2 & 7 & 2 & 6\tabularnewline
 & None & 14 & 9 & 14 & 8 & 14 & 9\tabularnewline
\hline 
\end{tabular}
\par\end{centering}
\caption{Statistical comparison of the number of unique words in each qualitative
code as judged by different coders across a subset of the data (the
top 16 words before and 15 words after).\label{tab:coders}}
\end{table}

\dataavailability{Data supporting the findings are contained within
the article and its supplementary material.}

\authorcontribution{MOA conceived the project and its evaluation,
performed the analysis, and wrote the paper. ND and SB assisted with
the analysis.}

\competinginterests{The authors declare that they have no conflict
of interest.}
\begin{acknowledgements}
We thank the researchers (Alice Giroul, Christopher Chen, Emma Davies,
Jesse Coburn, Joe Eggington, Luca Franci, Oleg Shebanits) and undergraduate
ambassadors (Avishan Shahryari, Cheng Yeen Pak, Christopher Comiskey
Erazo, Habibah Khanom, Safiya Merali, Yinyi Liu) who helped deliver
the exhibit along with all the staff at the Science Museum (including
Becky Carlyle, Imogen Small, Sevinc Kisacik). This project has been
supported by QMUL Centre for Public Engagement Large 2016 and Small
2019 Awards, an EGU Public Engagement Grant 2017, and STFC Public
Engagement Spark Award ST/R001456/1. M.O. Archer holds a UKRI (STFC
/ EPSRC) Stephen Hawking Fellowship EP/T01735X/1.
\end{acknowledgements}
\bibliographystyle{copernicus}
\bibliography{soundscape}

\begin{thebibliography}{40}
\providecommand{\natexlab}[1]{#1}
\providecommand{\url}[1]{{\tt #1}}
\providecommand{\urlprefix}{URL }
\expandafter\ifx\csname urlstyle\endcsname\relax
  \providecommand{\doi}[1]{https://doi.org/\discretionary{}{}{}#1}\else
  \providecommand{\doi}{https://doi.org/\discretionary{}{}{}\begingroup
  \urlstyle{rm}\Url}\fi

\bibitem[{3KQ and {Collingwood Environmental
  Planning}(2015)}]{spaceweatherdialogue15}
3KQ and {Collingwood Environmental Planning}: Space weather public dialogue,
  Tech. rep., Sciencewise, Science and Technology Facilities Council, RAL
  Space, Natural Environment Research Council, National Grid, Lloyd's of
  London,
  \urlprefix\url{https://www.ralspace.stfc.ac.uk/Pages/SWPDFinalReportWEB.pdf},
  2015.

\bibitem[{Agresti(2007)}]{agresti07}
Agresti, A.: An Introduction to Categorical Data Analysis (Second Edition),
  Wiley Series in Probability and Statistics, John Wiley \& Sons, Inc.,
  Hoboken, New Jersey, United States, \doi{10.1002/0470114754}, 2007.

\bibitem[{Alexander et~al.(2011)Alexander, Gilbert, Landi, Simoni, Zurbuchen,
  and Roberts}]{alexander11}
Alexander, R.~L., Gilbert, J.~A., Landi, E., Simoni, M., Zurbuchen, T.~H., and
  Roberts, D.~A.: Audification as a diagnostic tool for exploratory
  heliospheric data analysis, in: The 17th International Conference on Auditory
  Display, 2011.

\bibitem[{Alexander et~al.(2014)Alexander, O'{M}odhrain, Roberts, Gilbert, and
  Zurbuchen}]{alexander14}
Alexander, R.~L., O'{M}odhrain, S., Roberts, D.~A., Gilbert, J.~A., and
  Zurbuchen, T.~H.: The bird's ear view of space physics: {A}udification as a
  tool for the spectral analysis of time series data, J. Geophys. Res. Space
  Phys., 119, 5259--5271, \doi{10.1002/2014JA020025}, 2014.

\bibitem[{Archer and {DeWitt}(2017)}]{archer17}
Archer, L. and {DeWitt}, J.: Understanding Young People's Science Aspirations:
  How students form ideas about `becoming a scientist', Routledge, London, UK,
  \doi{10.4324/9781315761077}, 2017.

\bibitem[{Archer(2020{\natexlab{a}})}]{archer20entaudio}
Archer, M.~O.: In space no-one can hear you scream…or can they?, {ENT} and
  Audiology News, 28,
  \urlprefix\url{https://www.entandaudiologynews.com/features/audiology-features/post/in-space-no-one-can-hear-you-scream-or-can-they},
  accessed: Nov 2020, 2020{\natexlab{a}}.

\bibitem[{Archer(2020{\natexlab{b}})}]{archer_ssfx}
Archer, M.~O.: {Space} {Sound} {Effects} {Short} {Film} {Festival}:using the
  film festival model to inspire creative art-science and reach new audiences,
  Geosci. Commun., 3, 147--166, \doi{10.5194/gc-3-147-2020},
  2020{\natexlab{b}}.

\bibitem[{Archer et~al.(2018)Archer, Hartinger, Redmon, Angelopoulos, Walsh,
  and {Eltham Hill School Year 12 Physics students}}]{archer18}
Archer, M.~O., Hartinger, M.~D., Redmon, R., Angelopoulos, V., Walsh, B.~M.,
  and {Eltham Hill School Year 12 Physics students}: First results from
  sonification and exploratory citizen science of magnetospheric {ULF} waves:
  {L}ong-lasting decreasing-frequency poloidal field line resonances following
  geomagnetic storms, Space Weather, 16, 1753--1769,
  \doi{10.1029/2018SW001988}, 2018.

\bibitem[{Baixeries et~al.(2013)Baixeries, Elvev{\aa}g, and
  {Ferrer-i-Cancho}}]{baixeries13}
Baixeries, J., Elvev{\aa}g, B., and {Ferrer-i-Cancho}, R.: The Evolution of the
  Exponent of {Zipf}'s Law in Language Ontogeny, PLoS ONE, 8, e53\,227,
  \doi{10.1371/journal.pone.0053227}, 2013.

\bibitem[{Baumjohann and Treumann(2012)}]{baumjohann}
Baumjohann, W. and Treumann, R.: Imperial College Press, London, UK,
  \doi{10.1142/P850}, 2012.

\bibitem[{{BERA}(2018)}]{bera18}
{BERA}: Ethical Guidelines for Educational Research, Tech. Rep. Fourth Edition,
  British Educational Research Association, London,
  \urlprefix\url{https://www.bera.ac.uk/researchers-resources/publications/ethical-guidelines-for-educational-research-2018},
  accessed: Apr 2020, 2018.

\bibitem[{Braun and Clarke(2006)}]{Braun2006}
Braun, V. and Clarke, V.: Using thematic analysis in psychology, Qualitative
  Research in Psychology, 3, 77--101, \doi{10.1191/1478088706qp063oa}, 2006.

\bibitem[{Caleon et~al.(2013)Caleon, Subramaniam, and Regaya}]{caleon13}
Caleon, I., Subramaniam, R., and Regaya, M. H.~P.: Revisiting the bell-jar
  demonstration, Physics Education, 48, 247--251,
  \doi{10.1088/0031-9120/48/2/247}, 2013.

\bibitem[{Clopper and Pearson(1934)}]{clopper34}
Clopper, C. and Pearson, E.~S.: The use of confidence or fiducial limits
  illustrated in the case of the binomial, Biometrika, 26, 404--413,
  \doi{10.1093/biomet/26.4.404}, 1934.

\bibitem[{Feder(2012)}]{feder12}
Feder, T.: Shhhh. Listen to the data, Physics World, 65, 20--22,
  \doi{10.1063/PT.3.1550}, 2012.

\bibitem[{Galliano(2018)}]{galliano18}
Galliano, D.: SEPnet Outreach \& Public Engagement 2017/18 Reporting, Tech.
  rep., South East Physics Network, unpublished internal document, 2018.

\bibitem[{Grand and Sardo(2017)}]{grand17}
Grand, A. and Sardo, A.~M.: What Works in the Field? {Evaluating} Informal
  Science Events, Front. Commun., 2, \doi{10.3389/fcomm.2017.00022}, 2017.

\bibitem[{Gurnett et~al.(2013)Gurnett, Kurth, Burlaga, and Ness}]{gurnett13}
Gurnett, D.~A., Kurth, W.~S., Burlaga, L.~F., and Ness, N.~F.: In Situ
  Observations of Interstellar Plasma with {Voyager} 1, Science, 341,
  1489--1492, \doi{10.1126/science.1241681}, 2013.

\bibitem[{Hochberg and Tamhane(1987)}]{hochberg}
Hochberg, Y. and Tamhane, A.~C.: Multiple Comparison Procedures, Wiley Series
  in Probability and Statistics, John Wiley \& Sons, Inc., Hoboken, New Jersey,
  United States, \doi{10.1002/9780470316672}, 1987.

\bibitem[{{Hooper-Green}(2004)}]{hoopergreen07}
{Hooper-Green}, E.: Measuring Learning Outcomes in Museums, Archives and
  Libraries: The Learning Impact Research Project ({LIRP}), International
  Journal of Heritage Studies, 10, 151--174,
  \doi{https://doi.org/10.1080/13527250410001692877}, 2004.

\bibitem[{Jensen(2014)}]{jensen14}
Jensen, E.: The problems with science communication evaluation, J. Sci. Commun,
  13, \doi{10.22323/2.13010304}, 2014.

\bibitem[{Keiling et~al.(2016)Keiling, Lee, and Nakariakov}]{Keiling2016}
Keiling, A., Lee, D.-H., and Nakariakov, V., eds.: Low-Frequency Waves in Space
  Plasmas, Geophysical Monograph Series, American Geophysical Union,
  \doi{10.1002/9781119055006}, 2016.

\bibitem[{King et~al.(2015)King, Steiner, Hobson, Robinson, and
  Clipson}]{king15}
King, H., Steiner, K., Hobson, M., Robinson, A., and Clipson, H.: Highlighting
  the value of evidence-based evaluation:pushing back on demands for `impact',
  J. Sci. Commun, 14, \doi{10.22323/2.14020202}, 2015.

\bibitem[{Kramer(1994)}]{kramer94}
Kramer, G.: An Introduction to Auditory Display, Auditory Display:
  Sonification, Audification, and Auditory Interfaces, Addison-Wesley, Reading,
  MA, 1994.

\bibitem[{Lamers and Cassinelli(1999)}]{lamers99}
Lamers, H. J. G. L.~M. and Cassinelli, J.~P.: Introduction to stellar winds,
  Cambridge University Press, Cambridge, UK, 1999.

\bibitem[{Massey(1951)}]{massey51}
Massey, Jr., F.~J.: The {Kolmogorov}-{Smirnov} Test for Goodness of Fit, J Am
  Stat Assoc, 46, 68--78, \doi{10.1080/01621459.1951.10500769}, 1951.

\bibitem[{Maxwell(2010)}]{maxwell10}
Maxwell, J.~A.: Using numbers in qualitative research, Qualitative Inquiry, 16,
  475--482, \doi{10.1177/1077800410364740}, 2010.

\bibitem[{Piantadosi(2014)}]{piantadosi14}
Piantadosi, S.~T.: Zipf’s word frequency law in natural language: A critical
  review and future directions, Psychon. Bull. Rev., 21, 1112--1130,
  \doi{10.3758/s13423-014-0585-6}, 2014.

\bibitem[{{Public Engagement with Research team}(2019)}]{oxford19}
{Public Engagement with Research team}: Little Book of Evaluation Tools:
  Curiosity Carnival, Tech. rep., University of Oxford,
  \urlprefix\url{https://www.mpls.ox.ac.uk/public-engagement/latest/little-book-of-evaluation-tools-curiosity-carnival},
  accessed: Aug 2020, 2019.

\bibitem[{Robinson and Dadson(1956)}]{robinson56}
Robinson, D.~W. and Dadson, R.~S.: A re-determination of the equal-loudness
  relations for pure tones, Biritsh Journal of Applied Physics, 7, 166,
  \urlprefix\url{http://stacks.iop.org/0508-3443/7/i=5/a=302}, 1956.

\bibitem[{Robson(2011)}]{Robson2011}
Robson, C.: Real World Research, John Wiley and Sons Ltd., Hoboken, New Jersey,
  USA, 2011.

\bibitem[{Roeser et~al.(2007)Roeser, Valente, and {Hosford-Dunn}}]{roeser}
Roeser, R., Valente, M., and {Hosford-Dunn}, H.: Audiology: Diagnosis, Thieme,
  New York, USA, 2007.

\bibitem[{Sandelowski(2001)}]{sandelowski01}
Sandelowski, M.: Real qualitative researchers do not count: {T}he use of
  numbers in qualitative research, Res. Nurs. Health, 24, 230--240,
  \doi{10.1002/nur.1025}, 2001.

\bibitem[{Sandelowski et~al.(2009)Sandelowski, Voils, and
  Knafl}]{sandelowski09}
Sandelowski, M., Voils, C.~I., and Knafl, G.: On quantizing, J Mix Methods
  Res., 3, 208--222, \doi{10.1177/1558689809334210}, 2009.

\bibitem[{{Science Museum Group}(2017)}]{smg17}
{Science Museum Group}: Inspiring futures: Strategic priorities 2017-2030,
  Tech. rep., Science Museum Group, London, UK,
  \urlprefix\url{https://learning.sciencemuseumgroup.org.uk/wp-content/uploads/2020/04/Inspiring-Futures-Strategic-Priorities-2017-2030.pdf},
  accessed: Aug 2020, 2017.

\bibitem[{{Science Museum Group}(2020)}]{smg20}
{Science Museum Group}: Engaging all audiences with science: Science capital
  and informal science learning, Tech. rep., Science Museum Group, London, UK,
  \urlprefix\url{https://learning.sciencemuseumgroup.org.uk/blog/engaging-all-audiences-with-science-science-capital-and-informal-science-learning/},
  accessed: Aug 2020, 2020.

\bibitem[{Silverman(2010)}]{Silverman2010}
Silverman, D.: Doing Qualitative Research: A Practical Handbook,, Sage
  Publications Ltd., Thousand Oaks, California, USA, 2010.

\bibitem[{Supper(2014)}]{supper14}
Supper, A.: Sublime frequencies: the construction of sublime listening
  experiences in the sonification of scientific data, Social Studies of
  Science, 44, 34--58, \doi{10.1177/0306312713496875}, 2014.

\bibitem[{Zipf(1935)}]{zipf35}
Zipf, G.~K.: The psycho-biology of language, Houghton Mifflin, Boston, MA, USA,
  1935.

\bibitem[{Zipf(1949)}]{zipf49}
Zipf, G.~K.: Human behavior and the principle of least effort, Addison-Wesley
  Press, Boston, MA, USA, 1949.

\end{thebibliography}

\end{document}